\documentclass[nobibnotes,preprintnumbers,aps,prl,superscriptaddress,showpacs,twocolumn,twoside,sort&compress]{revtex4-1}              %
\usepackage{graphicx}                                                                                                                %
\usepackage{titlesec}
\usepackage{multirow}
\usepackage{fancyheadings}                                                                                                           %
\begin{document}                                                                                                                     %
\preprint{{Vol.XXX (201X) ~~~~~~~~~~~~~~~~~~~~~~~~~~~~~~~~~~~~~~~~~~~~~~~~~~~~ {\it CSMAG`16}                                         ~~~~~~~~~~~~~~~~~~~~~~~~~~~~~~~~~~~~~~~~~~~~~~~~~~~~~~~~~~~~ No.X~~~~}}                                                                %
\vspace*{-0.3cm}                                                                                                                     %
\preprint{\rule{\textwidth}{0.5pt}}                                                                                                          \vspace*{0.3cm}                                                                                                                         %

\title{Superconductivity of niobium thin films in the BiOCl/Nb heterostructures.}

\author{D. Lotnyk}
\thanks{Dmytro Lotnyk; e-mail: dmytro.lotnyk@student.upjs.sk}
\affiliation{Institute of Physics, Faculty of Science, P. J.
\v{S}af\'{a}rik University, Park Angelinum 9, 041 54 Ko\v{s}ice,
Slovakia}
\author{V. Komanick\'{y}}
\affiliation{Institute of Physics, Faculty of Science, P. J.
\v{S}af\'{a}rik University, Park Angelinum 9, 041 54 Ko\v{s}ice,
Slovakia}
\author{V. Bunda}
\affiliation{Transcarpathian Institute of Arts, Voloshiv st. 37,
88000 Uzhgorod, Ukraine}
\author{A. Feher}
\affiliation{Institute of Physics, Faculty of Science, P. J.
\v{S}af\'{a}rik University, Park Angelinum 9, 041 54 Ko\v{s}ice,
Slovakia}

\begin{abstract}
In the current paper, electrical transport properties of 25~nm
thick Nb films sputtered on the photosensitive semiconductor BiOCl
were investigated in the8 temperature range 7.5~K~$\leq T
\leq$~8.5~K. The influence of green (532 nm) and red (640 nm)
laser excitations on resistive superconducting transitions of the
niobium thin films on a silicon glass and BiOCl single crystal
substrates were studied. The temperature dependences of the
resistivity for Nb are in good agreement with the McMillan model
which indicates the strong influence of the inverse proximity
effect induced by the interface. The increased influence of the
BiOCl/Nb interface under laser excitation corresponds to the
raising the ratio of the density of normal to superconductivity
carriers in the $T\rightarrow 0$ limit and this observation is in
agreement with the photoconductivity study of BiOCl single
crystals.
\end{abstract}

\pacs{74.25.F- 68.35.bg 74.45.+c}

\maketitle

%
\section{Introduction}

In the past decade, attention was paid to the light-induced effect
in superconductors~\cite{Fausti, Mitrano, Suda}. Under laser
excitation, it is possible to manipulate the surface density of
state, to change superconducting properties. Another possibility
to change the local density of states arises from proximity
induced effects~\cite{Gennes}. One can manipulate the
superconducting density of states at the interface using
photosensitive semiconductor/superconductor heterostructures.
Bismuth oxyhalides BiOX (X = F, Cl, Br and I) photosensitive
semiconductors that have attracted intensive attention due to
their characteristic photoelectrical properties and possible
technological applications \cite{Wu, Gondal}. Due to their layered
structure and facile fabrication by exfoliation~\cite{Geim},
single crystals with thicknesses less than 1~$\mu$m could be used
as a substrate for semiconductor/superconductor heterostructures.

The electrical transport properties of thin Nb films, with nominal
thicknesses of 25~nm, supported on Si glass or BiOCl single
crystal substrates were investigated in the temperature range
7.5~K~$\leq T \leq$~8.5~K with and without laser excitation. The
temperature dependences of resistivity for Nb on Si glass
overlapping with each other indicating an excellent stability of
interface. The resistive curves at the vicinity of superconducting
transition for Nb on BiOCl single crystal are in good agreement
with the McMillan model~\cite{mcmillan}, which indicates the
strong influence of inverse proximity effect induced by the
interface. The calculated depth of the interface is approximately
0.85~nm. The increased influence of the interface under the laser
excitation corresponds to the raising in $N_N(0)/N_S(0)$ ratio
($N_{N,S}(0)$ are the density of states in normal (N) and
superconducting (S) layers at $T$~=~0). It is in agreement with
the photoconductivity spectra of BiOCl single crystals obtained
in~\cite{BundaSAM}.
\section{Experimental details \label{ss:expdet}}
In this work, a magnetron sputtering technique was used to
generate Nb films with nominal thicknesses of 25~nm. Sputtering
conditions were: chamber pressure 3.7$\times$10$^{-10}$~Pa, argon
pressure 3$\times$10$^{-5}$~Pa, DC target power 270~W, substrate
temperature 22~$^{\circ}$C. The thickness of the sputtered films
was controlled by quartz crystal microbalance device. The films
were simultaneously deposited on commercially purchased Si glass
slides 20~mm x 20~mm x 0.5~mm and BiOCl single crystal 10~mm x
5~mm x 1~$\mu$m obtained by exfoliation method~\cite{Geim}. The
high-quality BiOCl single crystal that served as the source of the
thin films was synthesized using gas transfer methods briefly
described in~\cite{BundaSAM}.
\begin{figure}[t]
\includegraphics[width=0.9\columnwidth]{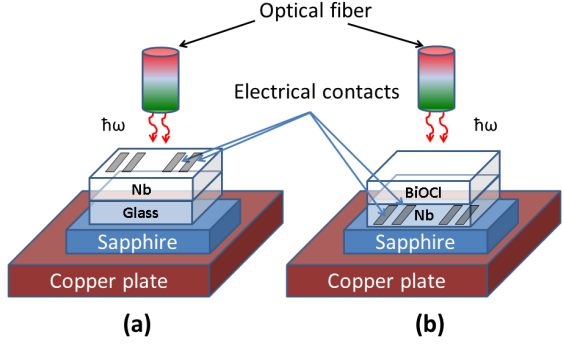}
\caption{Schematic representation of electric transport
measurements. a) 25 nm Nb sputtered on the Si glass and b) 25 nm
Nb sputtered on the BiOCl.} \label{fig:scheme}
\end{figure}

The electrical resistivity of the Nb thin films was obtained by
the four-point probe measurements of a direct current using
Keithley 6221 current source and Keithley 2182A nanovoltmeter.
Temperature control was performed using the commercial Physical
Property Measurement System (PPMS). Temperature dependences were
obtained in both cooling and heating modes at two temperature
rates 0.1~K/min and 0.3~K/min. Thermal contact was established by
installing sample on a massive Cu plate glued with the GE-Varnish
(see scheme in Fig.~\ref{fig:scheme}). Electric contacts were made
using the silver conducting paste with annealing at 80$^o$C during
10~min. The thin sapphire plate was used to avoid short circuit or
current leaking. Laser excitation with the power 15~mW/cm$^2$ was
applied on either Nb (Fig.~\ref{fig:scheme}a) or the
photosensitive semiconductor BiOCl (Fig.~\ref{fig:scheme}b). Laser
excitations with 532~nm (2.33~eV) and 640~nm (1.94~eV) were
obtained by two CNI laser devices with the maximum power 300~mW
and 200~mW, respectively. Output laser power was measured using
thermal power sensor (Ophir, 3A-P). Resistive curves were measured
at the current 1~$\mu$A.

\section{Results and discussion}
\begin{figure}[h!]
\includegraphics[width=0.9\columnwidth]{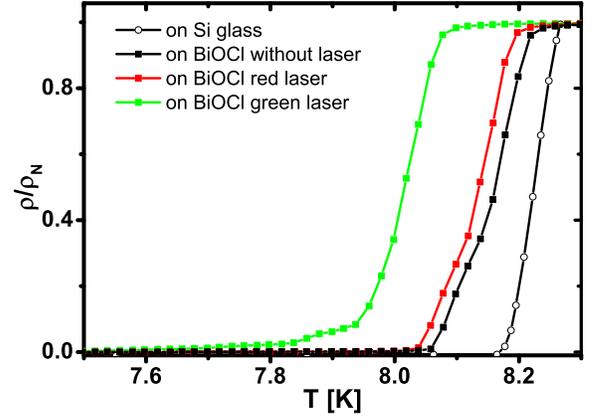}
\caption{Temperature dependences of normalized resistivity of
25~nm Nb thin films on Si glass (circles); on BiOCl without laser
excitation (black squares), under green (green squares) and red
(red squares) laser excitation.} \label{fig:rt}
\end{figure}

\begin{figure}[h!]
\includegraphics[width=0.9\columnwidth]{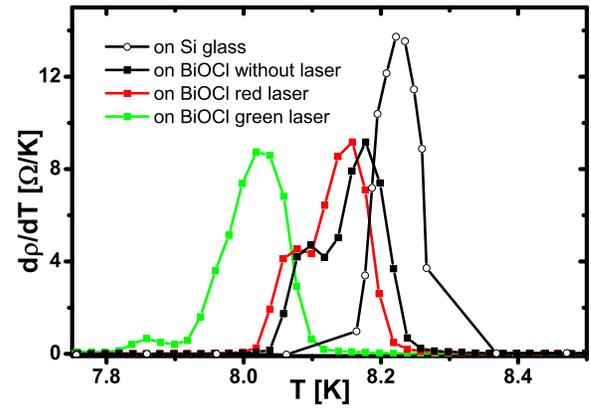}
\caption{Temperature dependences of $d\rho/dT$ for 25~nm Nb thin
films on Si glass (circles); on BiOCl without laser excitation
(black squares), under green (green squares) and red (red squares)
laser excitation.} \label{fig:drt}
\end{figure}

The experimental results of the superconducting transition for the
Nb films under various mounting and irradiation conditions are
given in Fig.~\ref{fig:rt}. It is noteworthy that the data for the
transitions were acquired with warming and cooling protocols at
two different rates, and the results  for a specific
mounting/irradiation conditions were independent of the various
measuring conditions. This observation indicates that a
high-quality thermal contact exists between the sample and the
cold plate of the PPMS, and consequently, the laser excitation
does not significantly perturb the temperature of the sample. Only
the curves measured at temperature rate 0.1~K/min in cooling mode
are plotted in Fig.~\ref{fig:rt}.
Critical temperatures were identified by the maximum value of
$d\rho/dT$ curves (fig.~\ref{fig:drt}), and the results are
tabulated in Table~\ref{tab:temp}. The width of superconducting
transitions was identified in the temperature range from deviation
from normal state behavior till the state $\rho\rightarrow 0$
(Fig.~\ref{fig:rt}). The thin film sputtered on Si glass has a
critical temperature 8.2~K along with the width of superconducting
transition $\Delta T_c\approx$ 0.2~K and the residual resistance
ratio (RRR) $\sim$~4.3 indicates a high quality of the thin film.
All measured curves, both without and with laser excitation,
perfectly match each other, and this result is a consequence of an
interface with excellent stability.

The superconducting transitionss of Nb on the BiOCl substrate
shift to the lower temperatures with increasing of the laser
excitation energy. The residual resistance ratio measured without
laser excitation 3.8, which is slightly lower compared to the Si
glass substrate. The higher width of the superconducting
transition $\Delta T_c$ and the presence of the second,
low-temperature peak $d\rho/dT$ curves could indicate the phase
separation in Nb thin films. With and without irradiation, the RRR
for Nb on BiOCl is the same, so the defect concentration is
constant during all measurements. On the other hand, the width of
the superconducting transition $\Delta T_c$ usually indicates the
impureness of the sample and this parameter increased sufficiently
under laser excitation.
\begin{table}
\begin{center}
\begin{tabular}{|c|c|c|c|c|}
\hline \multirow{2}{*}{Conditions} & \multicolumn{2}{|c|}{$T_c$,
K}& \multicolumn{2}{|c|}{$\Delta T_c$, K}
\\\cline{2-5}  & Si glass &
BiOCl & Si glass & BiOCl
\\\hline Without laser & 8.22 & 8.17 & 0.2 & 0.24
\\\hline Red light & 8.22& 8.15 & 0.2 & 0.28
\\\hline Green light & 8.22 & 8.02 & 0.2 & 0.35
\\\hline
\end{tabular}
\end{center}
\caption{Critical temperatures $T_c$ and the width of
superconducting transitions $\Delta T_c$ of 25~nm thin films on Si
glass and on the BiOCl substrates.} \label{tab:temp}
\end{table}

Such a behavior could be explained regarding inverse proximity
effect \cite{Delacour}. Despite the good quality of thin films,
which indicates in RRR~$\approx$~4, the pure Nb single crystals
have RRR up to 2600~\cite{Malang}. Consequently, the approximation
in dirty limit could be used. According to
Ashcroft~\cite{Ashcroft}, the electron mean free path was
estimated as $l=\frac{(r_s/a_0)^2}{\rho_{\mu}}\times
92\mathrm{\AA}$, where $\rho_{\mu}$ is a sheet resistance in
$\mu\Omega\cdot$cm and ratio $r_s/a_0$ for Nb is equals to 3.07.
The experimental value of sheet resistance is
$\rho_{\mu}$~=~3.09~$\mu\Omega\cdot$cm, which leads to the
$l\approx$~28~nm. Since the sample thickness $d$ is less than the
electron mean free path $l$, the former was used to evaluate the
coherence length in the dirty limit \cite{Ashcroft} as
$\xi$~=~0.852$\sqrt{\xi_0d}$. Assuming $\xi_0$~= 38~nm for bulk
Nb~\cite{Delacour}, then this analysis yields $\xi$~=~26~nm. This
value is very close to the thickness of our Nb film $d$~=~25~nm
and places our study in the two-dimensional limit. According to
the McMillan model \cite{mcmillan}, the inverse proximity effect
in a planar thin film geometry is an appropriate description of
our experimental conditions. Consequently, the critical
temperature $T_c$ is suppressed by the normal state interface with
the thickness $d_N$ described by the equation:
\begin{equation}
\centering
 T_c=T_{c0}\left(\frac{3.56T_D}{T_{c0}\pi}\right)^{-\alpha/d}, \label{eq:suppr}
\end{equation}
where $T_D$~=~277~K is the Nb Debye temperature, $T_{c0}$~=~9.22~K
is the Nb bulk critical temperature~\cite{Delacour} and
$\alpha~=~d_NN_N(0)/N_S(0)$. Considering a $N_N(0)/N_S(0)$ without
laser excitation, the calculated values of $d_N$ estimated as
0.85~nm. Under the laser excitation, the only parameter in
eq.~(\ref{eq:suppr}) being changed is $\alpha$. Consequently,
increasing of the $N_N(0)/N_S(0)$ ratio leads to the suppression
of the superconductivity. Changes for $N_N(0)/N_S(0)$ ratio are
1.02 under the red light excitation and 1.15 under the green light
excitation. Such behavior is in good agreement with
photoconductivity spectra obtained in~\cite{BundaSAM} for BiOCl
single crystals.

\section{Conclusions}
The electrical transport properties of 25~nm Nb thin films were
investigated in a temperature range 7.5~K~$\leq T \leq$~8.5~K on
Si glass and BiOCl substrates with and without laser excitation.
Resistive curves with and without laser excitation for Nb on Si
glass matched with each other which indicates a good stability of
interface and parameters in eq.~(\ref{eq:suppr}). The temperature
dependences of resistivity of Nb on BiOCl single crystal are in
good agreement with the McMillan model which indicates the strong
influence of inverse proximity effect induced by the interface.
The calculated depth of the normal state interface is
$d_N\approx$~0.85~nm. The increased influence of the interface
under the laser excitation corresponds to the enlarging in
$N_N(0)/N_S(0)$ ratio which is in agreement with the
photoconductivity study of BiOCl single crystals.
\section{Acknowledgements}
This work was supported by the ERDF EU (European Union European
regional development fond) grant, under the contract No. ITMS
26220120005, ITMS 26220220186, APVV 0605-14 and VEGA 1-0409-15. We
thank prof.~M. W. Meisel for the fruitful discussions and comments
that improved the manuscript.

\end{document}